\documentclass[eqsecnum,nofootinbib,floatfix]{revtex4} 
\usepackage{latexsym}
\usepackage{graphicx}
\usepackage{xspace}
\usepackage{psfrag}
\usepackage{amssymb}

\newcommand \VD {V_\Diamond}
\newcommand \nd {N_\Diamond}
\newcommand \be {\begin{displaymath}}
\newcommand \ee {\end{displaymath}}
\newcommand \bne {\begin{equation}}
\newcommand \ene {\end{equation}}
\newcommand \implies {\Rightarrow}
\newcommand{\gtsim}{\,\mbox{\raisebox{0.3ex}{$>$}\hspace{-0.8em}\raisebox{-0.7ex}{$\sim$}}\,}

\newcommand \ds {de Sitter\xspace}

\newcommand{\eprint}[1]{$\langle$arXiv: #1$\rangle$}

\begin{document}

\title{Indications of deSitter Spacetime from Classical Sequential Growth Dynamics of Causal Sets}

\author{Maqbool Ahmed}
\affiliation{Centre for Advanced Mathematics and Physics, 
Campus of College of E\&ME\\
National University of Sciences and Technology, \\
Peshawar Road, Rawalpindi, 46000, Pakistan}

\author{David Rideout}
\affiliation{Perimeter Institute for Theoretical Physics \\ 31
Caroline Street North, Waterloo, Ontario N2L 2Y5, Canada}

\date{6 November 2009}

\begin{abstract}

  A large class of the dynamical laws for causal sets described by a
  classical process of sequential growth yield a cyclic universe, whose
  cycles of expansion and contraction are punctuated by single `origin
  elements' of the causal set.  We present evidence that the effective dynamics of the immediate
  future of one of these origin elements, within the context of the
  sequential growth dynamics, 
yields an initial period of \ds-like exponential expansion,
 and argue that the resulting picture has many attractive features as
 a model of the early universe, with the potential to solve some of the standard
model puzzles without any fine tuning.

\end{abstract}

\maketitle
\tableofcontents

\section{Introduction}

Cosmology is the natural playground for theories of quantum gravity for many
reasons.  The most obvious is the fact that all quantum gravity theories are
formulated at such high energies (or small distances) that it is
practically impossible for an earth based laboratory or accelerator
to test them.  However, the early universe can provide such a
laboratory.  In addition cosmology is one of the most natural applications of general
relativity,
which is exactly what a theory of quantum gravity hopes to ``unify" with quantum
theory. On the other hand the standard model of cosmology~\cite{weinberg,mtw}
also suffers from some problems/puzzles~\cite{peebles} that
warrant extensions to it. Most of these have their origins in our
lack of understanding of the physics of very high energies and thus
a successful theory of quantum gravity should be able to address
these problems. 
Quantum gravity has recently began to shed some light on the puzzles of
cosmology, such as resolution of the cosmic singularity \cite{lqc, causetcosmo, turok},
providing some hints as to alternatives to inflation \cite{brandenberger},
and providing potential explanations of density
perturbations \cite{curv_perturbations}.
Additionally the prediction of a non-zero cosmological constant arises
naturally from the discreteness expected from quantum gravity
\cite{sor90}.
Thus quantum gravity is beginning to show some promise in resolving some of
the paradoxes of the standard model.  However, they are far from any
consensus in this regard, so cosmologists generally look for alternative explanations.

The most common path taken 
is an extension via particle physics that
introduces a scalar field with very special properties in the early
universe. If the potential of the field satisfies certain conditions
it can be arranged that the universe undergoes rapid expansion in
a ``\ds-like'' phase. This rapid expansion, called the inflationary
era or inflation,  gets rid of many of the standard model
puzzles/problems such as the so-called ``horizon problem", ``the
flatness puzzle", ``the monopole problem", etc. For a more
complete discussion see references~\cite{guth,earlyUni}. Despite the
fact that this scenario has ``problems" of its own, as has been
pointed out by many authors~\cite{earman,brandenberger}, the resulting
features are very attractive and hard to ignore. On the other hand
the lack of any competing model leaves a scientific void that needs
to be filled if the case for or against inflation has to be decided.

Fortunately Causal Set theory has reached a stage in its evolution
where some predictions about cosmology
have come out~\cite{sor90,everpLambda}. Based on a mixture of ``classical
dynamics" (referred to as the {\it classical sequential
growth}~\cite{cosacc} or {\it CSG models}) and some expectations
about ``quantum dynamics", Causal Set theory predicts fluctuations
in the cosmological term around a mean value.  If this mean value
is taken to be zero, the fluctuations are of the right magnitude to
explain the present observations. Computer simulations of the
behavior of the universe with such a fluctuating cosmological term
have shown that the energy density in $\Lambda$ follows the total
energy density of the universe and is roughly of the same order. This is the
first time that such testable predictions have come out of a
fundamental theory of quantum gravity.

Causal Set theory is also different from most
other quantum gravity theories in the
sense that it assumes fundamental discreteness. There have been many
arguments for discreteness at the most basic level, but all of them
have either been merely philosophical, or regarded only as methods of regulating
the 
infinities and singularities in particle physics and general
relativity. In the absence of real predictions it has always been
difficult to see if these considerations have more than a
philosophical value. Prediction about fluctuations in $\Lambda$, however,
draws life directly from fundamental discreteness, and it is but
natural to wonder if Causal Set theory has more to say about
cosmology. In this paper we will present evidence that many of
the CSG models produce a \ds-like early
universe, and thus may prove helpful towards solving another puzzle
--- why is the universe so large when it is not so old?

A general question which naturally arises in Causal Set theory is whether
causal sets which are well approximated by continua arise dynamically.  It
has been shown that the sequential growth models possess continuum limits, as
$N\to\infty$ and $p\to 0$, however the resulting continua look nothing like
spacetime manifolds of dimension $>1$~\cite{brightwell_georgiou}.  However it may still be the case that
something resembling a spacetime arises at finite $p$.  We consider this
latter question here.

This paper is organized as follows.  In section \ref{causet.sec} we briefly
describe that portion of Causal Set theory which is relevant to the current
work.  In section \ref{originary_perc.sec} we describe the behavior of the
``originary percolation'' dynamics, which arises as an effective dynamics of
the ``early universe'' of CSG models.  Then in section \ref{desitter.sec} we
compute the spacetime volume of `Alexandrov neighborhoods' (`causal
diamonds') in \ds space of arbitrary (integer) dimension.  In section
\ref{simulation.sec} we describe the particular simulation we perform, with
results in section \ref{results.sec}, and wrap up with some concluding
remarks in section \ref{conclusions.sec}.

\section{Causal Sets}
\label{causet.sec}
A causal set, or `causet' for short, is a locally finite partially ordered
set, whose elements can be thought of as irreducible\footnote{It may be
  necessary to add matter degrees of freedom to the causet elements, and at
  some stage it may be important to `coarse grain' the causet so that a single
  element may stand in for many, but for the moment we can think of the
  elements as not containing any internal information.}  `atoms of spacetime'.
A partially ordered set $C$ consists of a `ground set', which one generally
labels with integers from 0 to $N-1$ ($N$ can be infinite), along with a
binary relation $\prec$ which is irreflexive ($x \nprec x$) and transitive ($x
\prec y \prec z \implies x \prec z$).  Local finiteness is the condition that
every \emph{order interval} (or simply \emph{interval}) $[x,y] = \{ y | x
\prec y \prec z \}$ $\forall x, z
\in C$ has finite cardinality.

\subsection{Kinematics}
\label{kinematics.sec}
The connection to macroscopic spacetime arises via the notion of a
``sprinkling'', in which one selects events of a spacetime at random by a
Poisson process, identifies them with causal set elements, and then deduces a
partial ordering among the elements from the causal structure of the
spacetime.  One regards a continuum spacetime as being a good approximation to
an underlying causal set if that causal set is likely to have arisen from a
sprinkling into that spacetime.  For an extensive review of the causal set
program, see \cite{joe_review, valdivia, orig_prl, thooft}.

The connection between familiar concepts from continuum geometry and their
discrete counterparts on the causal set is the domain of causal set
kinematics.  We will make extensive use of two results in this regard.

The first is stated as a definition at this stage of the theory's development,
and is implicit in the description in the previous paragraph of the
correspondence of the causal set with the continuum.  In order for a causal
set to be likely to arise from a sprinkling, it must be the case that the
number of elements sprinkled into any region of spacetime with volume $V$ is
Poisson distributed, with a mean of $V$.  This Poisson fluctuation in the
correspondence between spacetime volume and number of elements plays a crucial
role in the prediction of a fluctuating cosmological constant \cite{sor90}.

The second result relates to the correspondence between the length of chains
and proper time.  A \emph{chain} is a subset of a causal set for which each
pair of elements is related.  The \emph{length} $L$ of a chain is the number
of elements in the chain minus 1.  In Minkowski space of any dimension, it has
been proven that the length of the longest chain between any pair of elements
is proportional to the proper time between the events at which they are
sprinkled, in the limit of infinite sprinkling density \cite{boxspace,ruth}.
In \cite{bachmat} the proportionality is claimed to hold for any spacetime.
Following \cite{ruth}, we define $m$ to be the constant of proportionality, so that
\bne
\tau = m L \;.
\label{tau_L.eqn}
\ene

\subsection{Dynamics}
\label{dynamics}
There are a number of approaches to constructing a dynamical law for causal
sets.  Perhaps the most developed to date is the classical sequential growth
model \cite{cosacc, d_thesis, becomming} mentioned in the Introduction.  It
describes the causal set as growing via a sort of ``cosmological accretion''
process, in which elements of the causet arise one at a time, each selecting
some subset of the causal set to be its past.
The process of growth in the model is stochastic; each newborn element selects
a ``precursor set'' at random, with probabilities which satisfy a discrete
analog of general covariance and a causality condition akin to that used to
derive the Bell inequalities.
This randomness is regarded as fundamental, and
yet purely classical in nature, because it does not allow for any quantum
interference among alternative outcomes.  Given the classical nature of the
probability distribution, the dynamics is incomplete, but can be seen as a
stepping stone toward formulating a fully quantum process, which could then
be regarded as a
generalization of classical probability theory.
Although the dynamically generated causal sets do not lead to orders which are
readily approximated by smooth spacetime manifolds, they do have a number of
striking cosmological features, which we explore further in this paper.

The sequential growth dynamics is described to take place in ``stages'', though
it is important to emphasize 
that the discrete general covariance condition enforces that this ordering in
which the causet elements arise is ``pure gauge'' --- it has no effect on the
probability of forming a particular (order equivalence class of) causal set.
At stage $n$ the
causal set has $n$ elements ``so far'', and the task is to select a precursor
set for the new element which arises in this stage.
The probabilities of the CSG model derive from a sequence of
nonnegative ``coupling constants'' $(t_n), n\geq 0$.  With these weights,
the probability for selecting a precursor set $S$ is proportional to
$t_{|S|}$.\footnote{The definition of precursor set used here differs from
  that in \cite{cosacc, d_thesis}.  There only precursor sets which contained
  their own past were included.  This definition allows for a much simpler
  expression of the transition probabilities (e.g.\ as given in
  \cite{brightwell_georgiou}), which better reveals the physical meaning of
  the $t_n$.}
Thus the probability to choose a particular set $S$ is
\be
Pr(S) = \frac{t_{|S|}}{\sum_{i=0}^n {n \choose i} t_i} \;.
\ee
Once a precursor is chosen, to be to the past of the new element,
all the relations implied by transitivity are included as well.  Thus it is
the ``past closure'' of $S$ which forms the past of the newly generated element.

The particular sequence
$t_n = t^n$, for a single non-negative real number $t$, gives rise to a
dynamics called \emph{transitive percolation} \cite{cosacc}.  This sequence
plays an important role, as we will see in a moment.  The rule for deciding
which elements to select for the past of a new element is particularly simple
for transitive percolation.  The newborn element simply considers each already
existing element in
turn, and selects it to be to its past with a fixed probability $p=t/(1+t)$.
It then adds to its past every element which precedes any of the originally
selected elements, to maintain transitivity.

\subsection{Cosmic Renormalization}

Consider an element of a causal set, called in the combinatorics literature a
`post', which is related to every other element of the causal set.  This would
resemble an initial or final singularity of a universe, in that the entirety
of the universe is causally related to it.  Now for any finite $p$, it has been
proven that a(n infinite) causet generated by transitive percolation almost
surely contains an infinite number of posts \cite{posts}.  It is the large
scale behavior of the universe subsequent to one of these posts which is the
subject of this paper.  We will present evidence that the period immediately
following a post is one of rapid expansion of spacetime volume with respect to
proper time.  Thus at the largest scales the causal sets generated by
transitive percolation resemble a bouncing universe, which periodically
undergoes collapse down to a final singularity and an ensuing re-expansion.

It has further been shown that a large class of CSG models, which includes
the sequence $t_n = (\alpha/\ln(n))^n$ for $n > 0$, $\alpha>\pi^2/3$ 
also lead to causets which
almost surely contain an infinite number of posts \cite{csg_posts}.
Now the presence of posts in the dynamical model suggests an interesting
possibility, as described in \cite{causetcosmo}, that the dynamics following a
post can be regarded as growing an entirely new universe, except with coupling
constants which are `renormalized' with respect to those of the previous era.
Much is now known about the flow of the coupling constants $(t_n)$ under this
`cosmic renormalization' \cite{cosrenorm, ash}, in particular that the transitive percolation
dynamics family $t_n=t^n$ forms a unique attractive fixed point.
The sequence $t_n = t^n/n!$ has also been studied in some detail
\cite{djamel_thesis,causetcosmo}.  There it is shown that the region
immediately subsequent to
a post behaves like transitive percolation, with a parameter $t$ which gets
driven toward zero under the cosmic renormalization ($t \to \sqrt{t/N}$ for
$N$ elements to the past of the current era's post (or ``origin element")).

\section{Originary Percolation and Random Trees}
\label{originary_perc.sec}
Note that the effective dynamics following a post comes with a caveat: each
element is required to be related to the post, by definition.  Therefore we
find an \emph{originary dynamics}, for which the possibility of being born
unrelated to any other element is excluded, and all remaining probabilities are
normalized correspondingly.  Thus the probabilities of an originary dynamics
are 
equal to those of an ordinary CSG model, conditioned on the event that the
newborn element connects to at least one other element.
(The originary dynamics is in fact one of the general class of solutions to
the covariance and causality conditions on sequential growth, described in
\cite{turtles}.)

\emph{Originary percolation} is the originary version of transitive
percolation.  As mentioned in section \ref{dynamics}, at each stage of the
growth process, the newborn element considers each existing element $x$ in turn, and
selects $x$ to be in its past with probability $p$.  In order to maintain
transitivity of the order, if it chooses $x$ for part of its past, it
includes all ancestors of $x$ as well. 
In the event that no element $x$ is selected in this process, it simply `tries
again', so as to maintain the condition of originarity.  At stage $n$ (meaning
that there are $n$ elements currently in the causet), the probability to
select a particular subset $S$ of existing elements is
\be
Pr(S) = \frac{p^{|S|} q^{n-|S|}}{1-q^n} \;,
\ee
where $q=1-p$, and the factor $1/(1-q^n)$ accounts for the originary
condition, which excludes the possibility of not connecting to anything (which
occurs with probability $q^n$).  
Once a set $S$ is
chosen, the past closure of $S$ becomes the past of the newborn element $x$.

\subsection{Random Tree Era}
\label{random_tree.sec}

For small values of $p$, the early universe of originary percolation, by
which we mean the structure of that portion of the causal set which is born
shortly after the origin element, forms a random tree, with high probability.
To see why this occurs, consider the probability that the selected precursor set
contains $m$ elements is given by
\be
Pr(|S|=m) = {n \choose m} \frac{p^{|S|} q^{n-|S|}}{1-q^n} \;.
\ee
For small $p$ this becomes vanishingly small for any $m>0$.  However the case
$m=0$ is excluded by the originarity condition, while it remains true that, as
long as $p \ll 1/n$, the transitions with $m=1$ will be much more likely than
any of the others.  
In these transitions one element is chosen at random from those already
present, with a uniform distribution.  This behaviour yields a simple model
of a random tree.
It persists until $n \sim 1/p$, at which point we get a percolation phase
transition, which heralds the end of the random tree era, and the beginning
of a phase of \ds-like expansion, which we describe below.
Note that the structure of this earliest random tree era of the universe is
independent of $p$, save in determining how long it lasts.

We can begin our study of the early universe of originary percolation by
studying this simple random tree process.  To get some feel for the initial
rate of
expansion of the universe, we ask
what is the expected number of elements which arise in `level
$t$', which we define to be those elements whose longest chain to the root
element is of length $t$.  With certainty, the first element appears in
level 0, and the second in level 1.  At stage $n$, the probability of
joining level $t$ is proportional to the number of elements in level $t-1$.
The same exact process has been studied in the combinatorics
literature, under the name ``random recursive trees''.  There similar
questions have been studied, such as the probability distribution of the level of an
element chosen uniformly at random from the tree
\cite{random_recursive_trees}.

\begin{figure}
\psfrag{t}{$t$}
\psfrag{Nt}{$N_t$}
\includegraphics{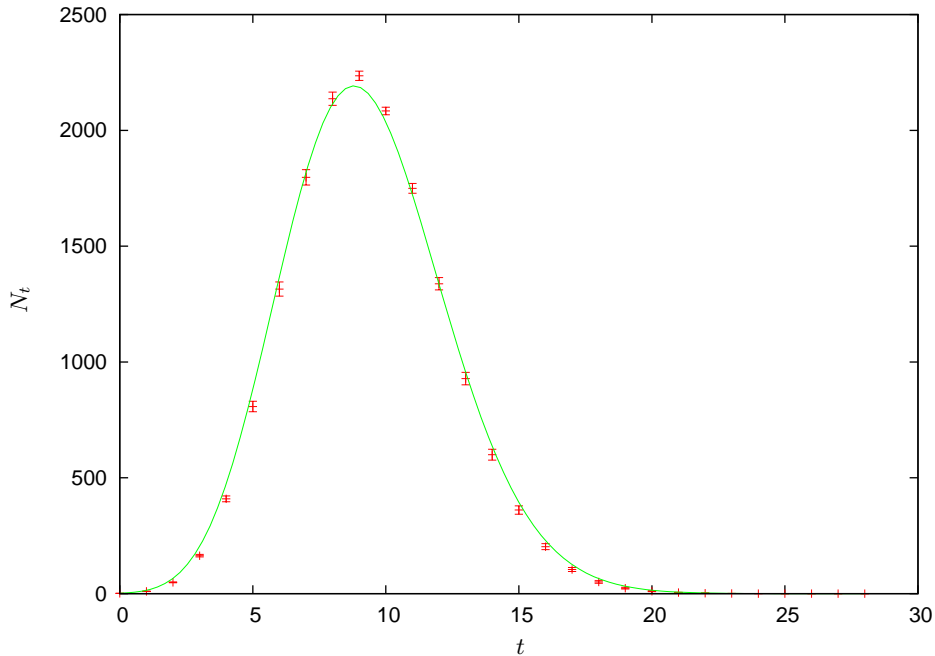}
\caption{Mean population of levels for an $N=16384$ element random tree.
  The mean and its error are measured from 100 samples of the random tree
  process.  The fitting function proportional to a Poisson distribution is
  shown.  Here $A=16687 \pm 85$ and $\lambda=9.306 \pm .022$.}
\label{poisson.fig}
\end{figure}

Despite the simple recursion obeyed by the `joining probability' above,
this problem is not easy to solve, e.g.\ because it involves an infinite
sequence of distributions.  Rather than analyze this problem in detail
here, we simply observe that, after forming a random tree with $N$
elements, the mean cardinality of level $t$ looks very much like a multiple
of a Poisson
distribution in $t$. 
Thus the mean number $N_t$ of
elements in level $t$ is very well fit by the function
\bne
N_t = \frac{A \lambda^t e^{-\lambda}}{t!} \;,
\label{poisson_fit}
\ene
where the normalization factor $A \gtsim N$ and $\lambda \sim \ln N$.  An
example is shown in figure \ref{poisson.fig}.
Despite the excellent fit of figure \ref{poisson.fig}, the relation cannot
be exact, for example because $N_t$ must be exactly zero for $t>N$, which
does not occur in (\ref{poisson_fit}).

The random tree era will continue until $n \sim 1/p$.  After this stage a
newborn element becomes as likely to choose more than one parent as not.
Thus we expect that the $N$ of the random tree era is $\sim 1/p$.
As far as the level population discussion goes, this is not the end of the
story, as
it is possible for such a `non-tree element' to select all its parents from
elements of the tree at an early level $t$, and thus itself to join an
early level, say one much earlier that the maximum of (\ref{poisson_fit}),
which is
$\sim \lambda \sim \ln(1/p)$.  Thus (\ref{poisson_fit}) provides only a lower bound on
the cardinality of early levels.

It is interesting to note that Gerhard 't Hooft predicted almost this exact
scenario in 1978, cf.\ figure 10 of \cite{thooft}.

\subsection{Originary Percolation}

To get a better handle on the initial rate of expansion, we perform
simulations of the full originary percolation dynamics.  This task is
greatly simplified through the use of the CausalSets toolkit within Cactus
framework \cite{cactus}.  All we need to do is write a `thorn' (module)
which counts the number of elements in each level, and counts the number of
elements and longest chains in each order interval, as explained in section
\ref{simulation.sec}.  The ability to generate causal sets via the
originary percolation dynamics is already provided within
the toolkit.

\begin{figure}[htbp]
\includegraphics[width=3.4cm]{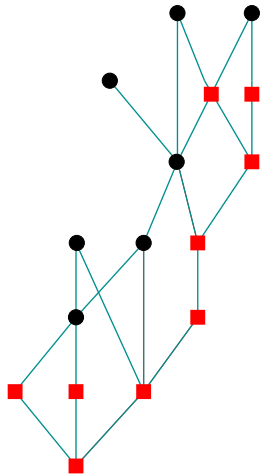}
\caption{A sample causal set generated by originary percolation with $N=16$,
  $p=0.2$.  Elements of the initial tree are shown by red squares.}
\label{16.2.fig}
\end{figure}
As an illustration, we show in figure \ref{16.2.fig} a small example causal
set generated by originary percolation with $N=16$,
$p=0.2$.  The past of any
element is the set of elements which can be reached from it by traversing the
lines (`links') downward.  The origin element / post is at the bottom.  The
red squares are elements which are part of the tree era.

\begin{figure}[htbp]
\psfrag{t}{$t$}
\psfrag{Nt}{$N_t$}
\includegraphics[width=\textwidth]{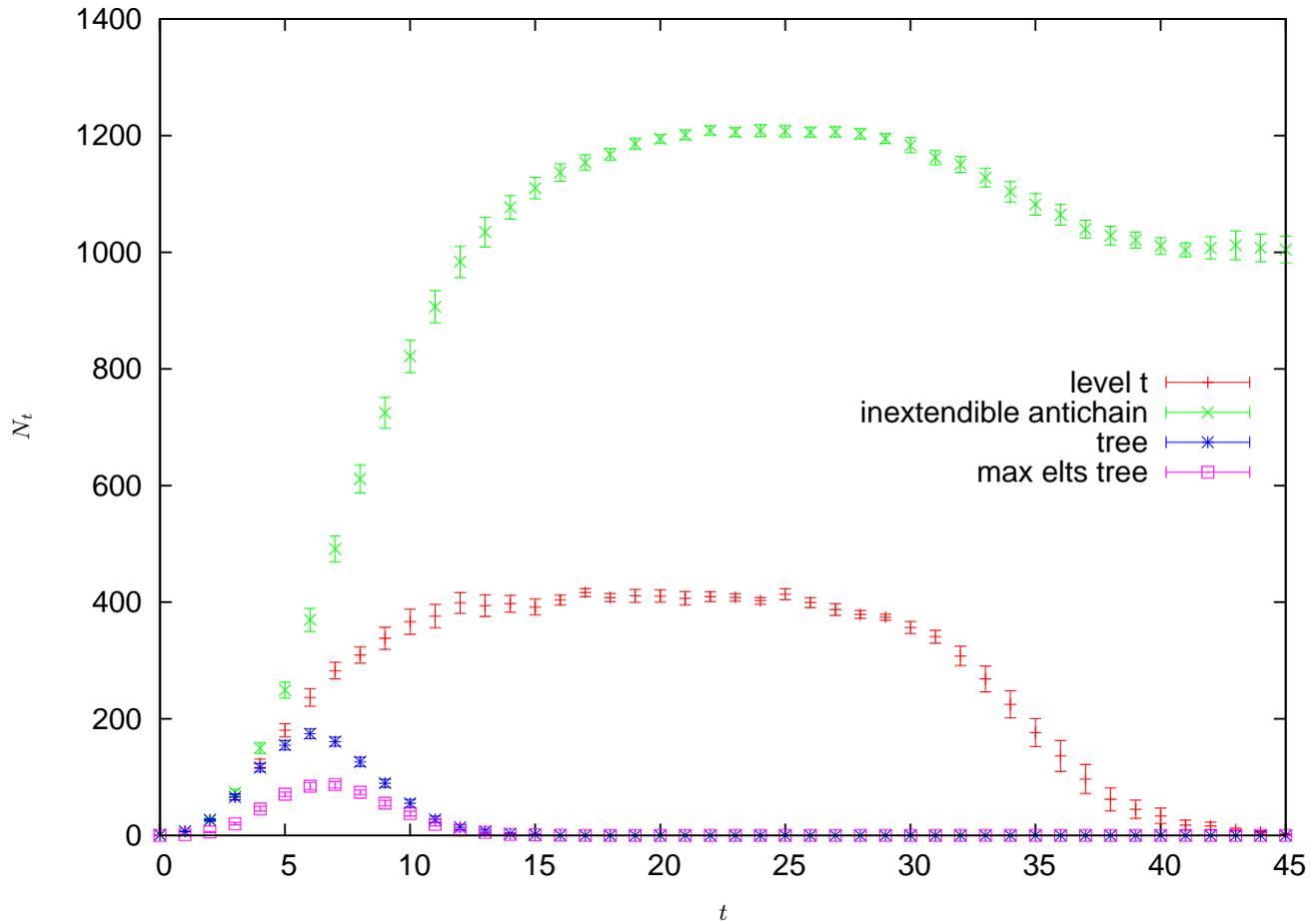}
\caption{Mean `spatial volume' of 10 originary percolated causal sets with
  $N=11585$, $p=.001$.  Shown is the cardinality of level $t$ in red, the
  cardinality of an inextendible antichain containing level $t$ in green, the
  number of elements of the initial tree at level $t$ in blue, and the number
  of maximal elements of the initial tree, at level $t$, in magenta.}
\label{spvol.fig}
\end{figure}
Results for originary percolation at $p=.001$, $N=11585$, are depicted in
figure \ref{spvol.fig}.
In
addition to the cardinality of each level mentioned above, we compute the
cardinality of a `foliation' of the causal set by inextendible antichains.
This is a more appropriate analogue to the (edgeless) spatial hypersurfaces of
General Relativity.
An \emph{antichain} is a subset of the causal set which contains no
relations.  An \emph{inextendible antichain} is one which is maximal in the
sense that no elements can be added to it while remaining an antichain,
i.e.\ every other element of the causal set is to the future or past of one
of its elements.  The inextendible antichains we employ here are defined as follows.
The level $t$ as defined above forms an antichain, but in general it will not
be inextendible.  We can extend it by adjoining the maximal elements (ones
which have no elements to their future) of that portion of the causal set
which is unrelated to any element of level $t$.  It is easy to show that
this will always yield an inextendible antichain.  Note that all of
the sub-causal set which is unrelated to the level $t$ antichain lives in
levels $<t$, for otherwise there would be a past directed chain
to some element of level $t$.  This fact motivates the choice of using the
maximal elements to form the inextendible antichain.\footnote{It turns out
  that this inextendible antichain is equivalent to the one which arises by
  taking the maximal elements of those whose level is $\leq t$.}

A final question that we consider before turning our attention to \ds
spacetime regards how the initial random tree sits within the larger
percolated causal set.  To this end we define an element to be within the
`tree era' if the order interval between it and the origin element is a chain.
In figure \ref{spvol.fig} 
we plot, in
addition to the cardinality of the antichains discussed above, the number
of elements in each layer that are part of the tree era.  
We see that if $N \gg 1/p$ then the initial tree sits in the very early part
of the percolated causal set.  The exponential expansion extends well
beyond the tree era, and thus the initial exponential growth of
(\ref{poisson_fit}) is indeed only a precursor to an ensuing exponential
growth involving a much larger portion of the causal set.

Before closing this section, it is important to note that the future of
every element of a percolated causal set is itself an instance of originary
percolation.  This is simply because percolation is completely homogeneous
--- the future of an element is the same (in probability) as that of any
other element.  However by discussing the future of an element $x$ one is
conditioning on each element being to the future of $x$, which is exactly
the condition of originary percolation.  Thus originary percolation
describes a homogeneous universe, for which the future of every element is
exponentially expanding.  This sounds a lot like \ds space.

\section{Volume of an Alexandrov Neighbourhood in deSitter Spacetime}
\label{desitter.sec}
For a spacetime respecting `the
cosmological principle', an exponential expansion  means
{\it the de Sitter spacetime}. If the
universe is described by something like a causal set, the
early universe region that we consider
is very young. 
It does not look like
a spacetime yet in the sense that 
it does not render itself easily to many of the familiar
concepts of the continuum. This is particularly clear
if one considers, for example, the random tree era.  
It is not possible to define the
notion of `spacelike distance' in a random tree as no two elements
have a common future \cite{spatial_dist}. Similarly it is difficult to see 
what curvature would mean in this case.  
On the other hand the notions of the length of the longest chain between two
causal set elements ($L$),
which is proportional to the proper time between the
two events ($\tau)$, and the number of causal set elements $N_\Diamond$
which are causally between two
given elements,\footnote{For an order interval $[x,y]$, we
  define $N_\Diamond = |\{z|x \prec z \prec y\}|+1$, where $|\cdot|$ indicates
  set cardinality.  The +1 allows $\nd = L$ for an interval which is a
  chain.}
which is proportional to the
volume of the Alexandrov neighborhood\footnote {The Alexandrov neighborhood of two events
is the overlap of the past of the futuremost event with
the future of the other.} $V_\Diamond$ formed by
the two elements,
is still defined.
We try to see if
$L$ and $\nd$ follow the same relationship as $\tau$ and $V_\Diamond$
in $D+1-$dimensional de Sitter spacetime.

\begin{figure}[h]
\scalebox{.6}{
\psfrag{t}{\LARGE $t$}
\psfrag{tr}{\Huge{$r$}}
\psfrag{R}{\LARGE $R$}
\psfrag{t1}{\LARGE $t_1$}
\psfrag{t2}{\LARGE $t_2$}
\includegraphics[width=7cm]{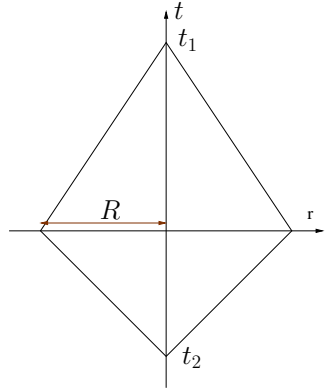}
}
\caption{An Alexandrov neighborhood (order interval) in \ds space.}
\label{desitter_interval}
\end{figure}

We use
\be
ds^2 = -dt^2 + e^{2t/\ell}(dr^2 + r^2 d\Omega_D^2)
\ee
as the $D+1-$dimensional de Sitter metric \cite{HawkingEllis}, where $\ell$ is
the radius of curvature
and all other symbols have their usual meaning.
As we have spherical symmetry in this case
we can represent the Alexandrov neighbourhood of two
events in $t$ and $r$ space as sketched in figure~\ref{desitter_interval}.
The spacetime volume of this region can be written as
\bne
\VD = \int_{-t_1}^0 dt e^{Dt/\ell} \int_0^{r_o} dr r^{D-1} \int d\Omega_D
+\int_0^{t_2} dt e^{Dt/\ell}\int_0^{r_i} dr r^{D-1} \int d\Omega_D \;.
 \label{int1}
\ene
As the light cones in de Sitter space follow $\dot{r} = \pm e^{-t/\ell}$
and we choose outgoing $r_o$ and ingoing $r_i$ radial coordinates 
such that $r_o(0) =r_i(0)= R$, we can write
$r_o = R+\ell(1-e^{-t/\ell})$ and
$r_i = R+\ell(e^{-t/\ell}-1)$. Using these we can write
(\ref{int1}) as
\bne \label{int2}
\VD = \mathcal{C}_D \ell^D \left[\int_{-t_1}^0 dt
      \left({{R+\ell} \over \ell}e^{t/\ell}-1  \right)^D
      +\int_0^{t_2} dt \left({{R-\ell} \over \ell}e^{t/\ell}+1  \right)^D \right]
\ene
where $\mathcal{C}_D$ is the volume of a $D$-dimensional unit ball.
Using $t_1 = \ell \ln{{\ell+R} \over \ell}$,
$t_2 = -\ell \ln{{\ell-R} \over \ell}$ and $t_1 + t_2 = \tau$,
we can simplify (\ref{int2}) to
\bne \label{oddD}
\VD = \mathcal{C}_D \ell^{D+1} \left[  \ln \cosh^2({\tau \over 2\ell})
+ \sum\limits_{i = 1}^D \frac{(-1)^{i+1}}{i} {D \choose i}
\left( \left(1+\tanh({\tau \over 2\ell})\right)^i+
\left(1-\tanh({\tau \over 2\ell})\right)^i -2 \right)  \right]
\ene
for $D$ odd and
\bne \label{evenD}
\VD =\mathcal{C}_D \ell^{D+1}\left[  {\tau \over \ell}
+ \sum\limits^D_{i = 1}{(-1)^i \over i} {D \choose i}
\left( \left(1+\tanh({\tau \over 2\ell})\right)^i-
\left(1-\tanh({\tau \over 2\ell})\right)^i\right)  \right]
\ene
for $D$ even.
One obvious case of interest is D = 3. Using the above
mentioned expressions and the fact that $\mathcal{C}_3 = 4\pi/3$
it turns out that
\be
\VD = {4\pi \over 3}\ell^4\left(\ln \cosh^2({\tau \over 2\ell})
- \tanh^2({\tau \over 2\ell})\right).
\ee
for a $4-$dimensional de Sitter spacetime.
It should be noted that $\VD \sim \tau^{D+1}$ for
${\tau \over \ell} \ll 1$ as every spacetime looks locally like
Minkowski space of the same dimension and $\sim \tau$ for
${\tau \over \ell} \gg 1$. For the $4-$dimensional
de Sitter Space  
$\VD = {\pi \over 24} \tau^4 + O(\tau^5)$
for $\tau \ll \ell$
  and $\approx 4\pi/3(\tau-\ln 4e)$
for $\tau \gg \ell$).

\section{Simulation details} 
\label{simulation.sec}

We want to compare the relationship between $\VD$
and $\tau$ given by equations
(\ref{oddD}) and (\ref{evenD}) with that 
produced by originary percolation between $N_\Diamond$ and $L$.
In a given simulation
we generate a causal set via originary percolation,
with a given  number of elements $N$ and the percolation parameter $p \in [0,1]$.
We then calculate the lengths of the longest chains $L$ between all pairs
of elements and the corresponding number of
elements $\nd$ that are connected to both of these elements
and lie causally between them. 
For this exercise the typical values of $N$ lie
between $1000$ and $50 000$ and of $p$ between
$.0001$ and $.03$.
The primary computational constraint is run time, as finding the length of
the longest chain in an interval involves an $O(N^2)$ algorithm, and there
are $O(N^2)$ intervals to check.

\begin{figure}[htbp]
\psfrag{L}{$L$}
\psfrag{Nd}{$\nd$}
\includegraphics[width=\textwidth]{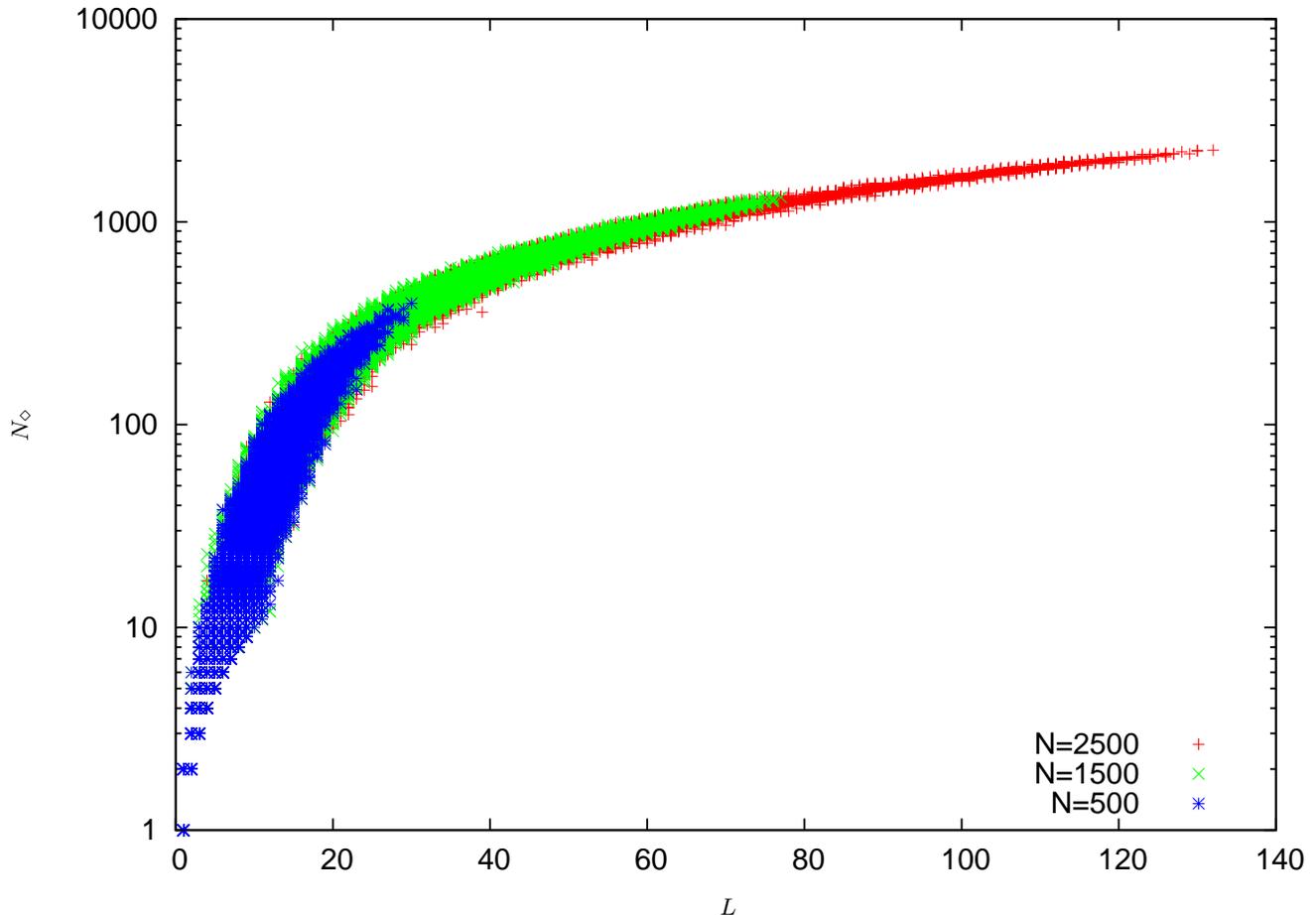}
\caption{The set of all pairs $(L, \nd)$ for each related pair of elements,
  in three causal sets.  Each causet 
  was generated with the same value of $p=.025$ but three different values of
  $N=$500, 1500, and 2500.
  (To make the figure size manageable, we plot only
  every $4^\mathrm{th}$ point for $N=500$, every $10^\mathrm{th}$ for
  $N=1500$, and every $200^\mathrm{th}$ point for $N=2500$.)
Note that the
  smaller data sets are a subset of the larger ones, and that at some point
  the maximum $\nd(L)$ no longer increases with $N$.}
\label{ordered_pairs}
\end{figure}

For a given causal set, we collect a large number of pairs of numbers
$(L, \nd)$, one for every related pair of elements in the causet.  The set of
such pairs for three causal sets is plotted in figure \ref{ordered_pairs}.
We wish to compare these data points with the functional forms (\ref{oddD})
and (\ref{evenD}), for some value of the dimension $D$.  If these causal sets
are exactly represented by \ds space, i.e.\ if they arose from a Poisson sprinkling
of a region of \ds of spatial dimension $D$, then one would expect the data
points to be scattered about the curve (\ref{oddD}) or (\ref{evenD}), with
Poisson fluctuations.  There are indications that, in spacetime dimensions
larger than 3, the fluctuations in the length of the longest chain in a
sprinkled interval of Minkowski space grows only logarithmically with $L$
\cite{llc}, so one might guess that we would see data points
distributed roughly uniformly above the curve (\ref{oddD}) or (\ref{evenD}).
However, for reasons we do not fully understand, it turns out
that the data points all seem to fall \emph{below} the curve
(\ref{oddD}) or (\ref{evenD}), such that the \emph{maximum} value of $\nd$
for a given $L$, for an appropriate range of values of $L$, gives an
excellent match to one of the functions (\ref{oddD}) or (\ref{evenD}).

It is important to notice that
almost all of the physics in this scenario is dictated by the choice
of $p$, as long as $N \gg 1/p$. 
This can be easily seen from figure~\ref{ordered_pairs}, by observing
that the data points for smaller $N$ are effectively a subset of those
for a larger value of $N$. Notice in particular that the maximum values of
$\nd$ for
the $N=1500$ causet are
the same as those for the $N=2500$ causet.
Thus, as long as $N$ is large enough to capture
the relevant region of exponential expansion, increasing $N$ further will
have no effect on the results of interest.\footnote{It is true that
  originary percolation for fixed $p$ and infinite $N$ contains every finite
  partial order as a suborder.  Thus somewhere in that infinite causet is an
  interval with height $L$ and cardinality arbitrarily large.  However, we do
  not send $N\to\infty$ for fixed $p$, we are only interested in the `early
  universe' of originary percolation, with $N$ no larger than say $1/p^3$.
  In such a regime the maximum $\nd$ is effectively independent of $N$.}
In particular this means that the dimension $D$ which gives the best fit,
for example, will only depend on $p$.
If $N$ is too small, 
on the other hand, then the
causal set is not large enough to `sample the region of interest', and we
will get poor results.  This is manifested in figure \ref{ordered_pairs} by
the fact that the maximum $\nd$ for $N=500$ are substantially smaller than
those for the larger causets.

The reader may be concerned that we use the maximum $\nd$ for a given value
of $L$, rather than the mean.  This is an indication that the percolated
causal set is not exactly manifoldlike.  However this is not too surprising,
as we already know that the CSG models do not have non-trivial spacetimes as
their continuum limits \cite{brightwell_georgiou}.  Another indication that
these are not quite manifoldlike is that at the smallest scales they are
trees, as explained in section \ref{random_tree.sec}, and thus one
dimensional (because the shortest intervals will always be chains).
This failure of the mean to give good results may be expected, in that it
gets contributions from all sorts of intervals,
including ones that might be `close to a boundary', such that they have small $\nd$ for large $L$.
In a sense we are considering only 
intervals as measured by observers which are stationary in
the cosmic rest frame, so that they can get the most elements for a given
proper time separation.

Since each causal set only provides a single maximum $\nd$ for each $L$, we
repeat the computation for a number of causal sets, and from these compute a
mean maximum $\nd$ with its error.
We then fit each
such data set with the expressions
given in equations~(\ref{oddD}) and (\ref{evenD}), with
$\tau$ replaced by $L/m$.  $\ell$ and
$m$ are used as fitting parameters.
For 
$D = 3$ the fitting expression looks like
\bne
\VD = {4\pi \over 3}\ell^4\left(\ln \cosh^2({\tau \over 2\ell})
- \tanh^2({\tau \over 2\ell})\right) \;.
\label{fitting_fn.eqn}
\ene
As mentioned above, at the smallest scales $L$ the causal set behaves like a
tree and is therefore, in the sense of order intervals, 0+1 dimensional.  At the largest scales the
intervals `see the infrared cutoff' $N$, and therefore are not expected to
give good results.  We thus only fit our data within a range of $L$ values,
as shown in table \ref{PvsD}.
Furthermore, since the error bars are much smaller for the small intervals
than for the large ones, fitting directly to the forms above would strongly
favor the small scales, and tend to ignore the data for larger scales.  We
handle this by fitting (the log of the maximum $\nd$) to the log of the functions above such as
(\ref{fitting_fn.eqn}), which has the effect of fitting to the relative error
in the maximum $\nd$.

At no point have
we ever mentioned any number for the dimension, in expressing the dynamics.  
Thus we have no idea what dimension of \ds space to expect from our results.
We therefore fit our data to every (spatial) dimension, usually from 1 to 9,
and take the one which fits best.

\section{Results}
\label{results.sec}

\begin{figure}[htbp]
\psfrag{data2}{\hspace{25mm}\footnotesize max $\nd$}
\psfrag{f2}{\hspace{18mm}\footnotesize 2+1 dS}
\psfrag{f3}{\hspace{18mm}\footnotesize 3+1 dS}
\psfrag{f4}{\hspace{18mm}\footnotesize 4+1 dS}
\psfrag{Ndiamond+1}{$\nd+1$}
\psfrag{L}{$L$}
\includegraphics[width=\textwidth]{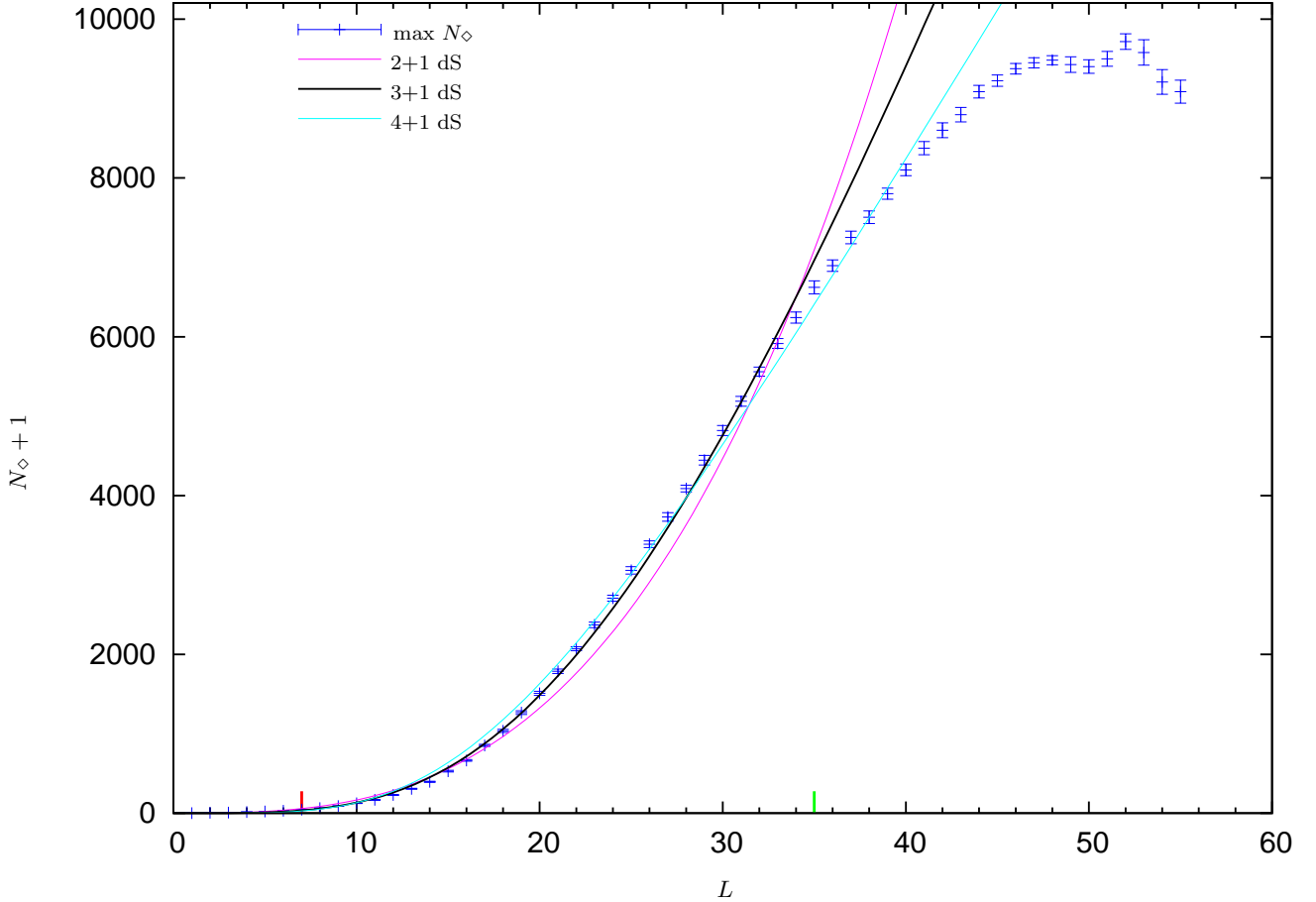}
\caption{A plot of the maximum values of $\nd$ as a function of $L$ for
  $p=0.001$ and $N=15000$, along with best fit curves for \ds space of three
  different dimensions. The vertical line on the left
  marks the end of the tree era,
  while the one on the right separates the points that `see' the finiteness
  of the causal set.  The overall best fit is achieved from the curve for 3+1
  dimensions, and is shown in black.}
\label{p0.001.fig}
\end{figure}

Figure~\ref{p0.001.fig} shows a typical behavior of the plot of the maximum
number of elements in an 
order interval and the corresponding
longest chains, for $N= 15000$ and $p =0.001$.
Interestingly enough, the best fit was achieved by the function for 3+1
dimensional \ds space, which is shown in black.  The best fits for the two
neighboring dimensions are also shown, to give some indication 
of the robustness of the dimension `measurement'.

The results for all our runs, for a variety of values of $p$, are summarized
in table~\ref{PvsD}.  
\begin{table}[htbp]
\begin{tabular}{|l|c|c|c|c|c|c|c|}
\hline
$p$&$n$&$D+1$&$\ell$&$m$& $\chi$&Fitting range in $L$&Number of runs \\
\hline
0.0001&50000&4&$8.7 \pm 1.8$&$2.105 \pm 0.028$&$2.32$&$9-25$&$6$\\
0.0002&30000&4&$6.81 \pm 0.72$&$1.926 \pm 0.023$&$3.07$&$8-27$&$3$\\
0.0005&20000&4&$7.81 \pm 0.57$&$1.787 \pm 0.022$&$5.59$&$7-32$&$10$\\
0.0008 & 15000 & 4 & $6.86 \pm 0.22$ & $1.749 \pm 0.019$ & 4.69 & $6-35$ & 4\\
0.001&15000&4&$6.20 \pm 0.12$&$1.710 \pm 0.013$&$4.97$&$7-35$&$23$\\
0.003&15000&4&$ 3.73\pm 0.013$&$1.483 \pm 0.009$&$5.12$&$6-100$&$20$\\
0.005&15000&4&$ 3.097\pm 0.009$&$1.388 \pm 0.009$&$4.69$&$5-150$&$20$\\
0.01&2000&3&$ 4.086 \pm 0.028$&$1.136 \pm 0.006$&$2.75$&$5-39$&$50$\\
0.03&1000&3&$ 2.331\pm 0.011$&$1.046 \pm 0.006$&$0.663$&$5-53$&$5$\\
\hline
\end{tabular}
\caption{Fitting parameters for some values of $p$.}
\label{PvsD}
\end{table}
All fits are performed with the gnuplot fit function.  For each value of $p$
we have considered, table \ref{PvsD} provides the value of $N$ we have used,
the best fit spacetime dimension, the best fit values for $\ell$ and $m$ with their
errors, $\chi = \sqrt{ \sum_L \frac{(N_{\Diamond,L} - \VD(L))^2}{(N_L-2)
    \:\sigma_{\nd,L}^2} }$
(where $N_L$ is the number of data points fit), the range of $L$ values we fit, and also the number of
causal sets generated.  All reported errors are as given by gnuplot.

As discussed in section \ref{kinematics.sec}, proper times are expected to be
related to length of the longest chain by (\ref{tau_L.eqn}).  If we assume
that the largest intervals of our causal sets do indeed behave like intervals
of \ds space, then the fits of table \ref{PvsD} serve as an alternate
measurement of $m$, in \ds spacetime of 3 and 4 dimensions.  It is
interesting to see that the values come out comparable to those for Minkowski
space, which fall between 1.77 and 2.62 \cite{ruth}.
The $\ell$ measurements indicate that we can grow a universe which is roughly
$2m\ell = 36$ elements `across'.

\begin{figure}[htbp]
\psfrag{data2}{\hspace{25mm}\footnotesize max $\nd$}
\psfrag{f2}{\hspace{18mm}\footnotesize 2+1 dS}
\psfrag{f3}{\hspace{18mm}\footnotesize 3+1 dS}
\psfrag{f4}{\hspace{18mm}\footnotesize 4+1 dS}
\psfrag{Ndiamond+1}{$\nd+1$}
\psfrag{L}{$L$}
\includegraphics[width=\textwidth]{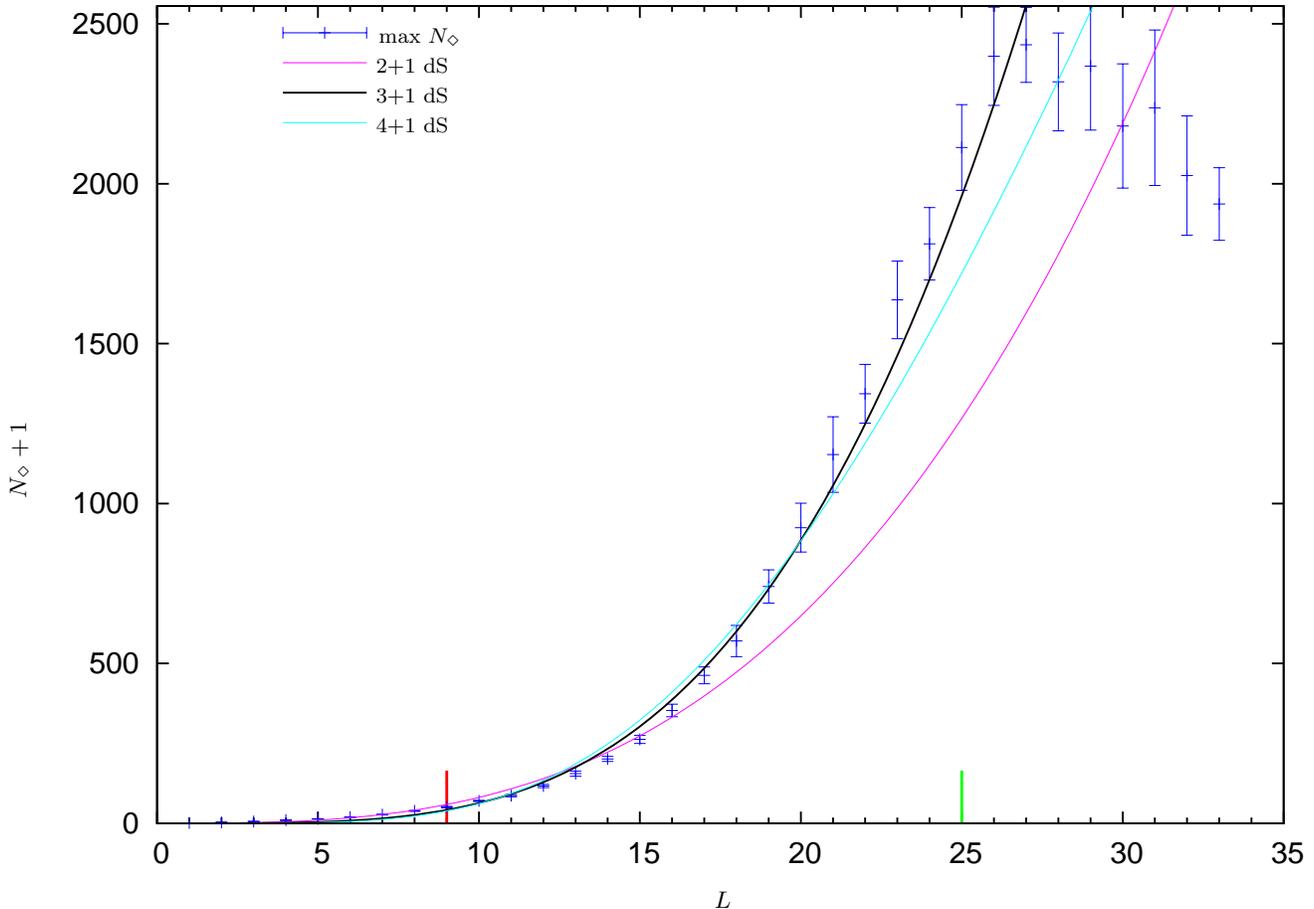}
\caption{A plot of the maximum values of $\nd$ as a function of $L$ for
  $p=0.0001$ and $N=50000$, along with best fit curves for \ds space of three
  different dimensions.
  The overall best fit is achieved from the curve for 3+1 dimensions.
}
\label{p0.0001.fig}
\end{figure}

\begin{figure}[htbp]
\psfrag{data2}{\hspace{25mm}\footnotesize max $\nd$}
\psfrag{f2}{\hspace{18mm}\footnotesize 2+1 dS}
\psfrag{f3}{\hspace{18mm}\footnotesize 3+1 dS}
\psfrag{f4}{\hspace{18mm}\footnotesize 4+1 dS}
\psfrag{Ndiamond+1}{$\nd+1$}
\psfrag{L}{$L$}
\includegraphics[width=\textwidth]{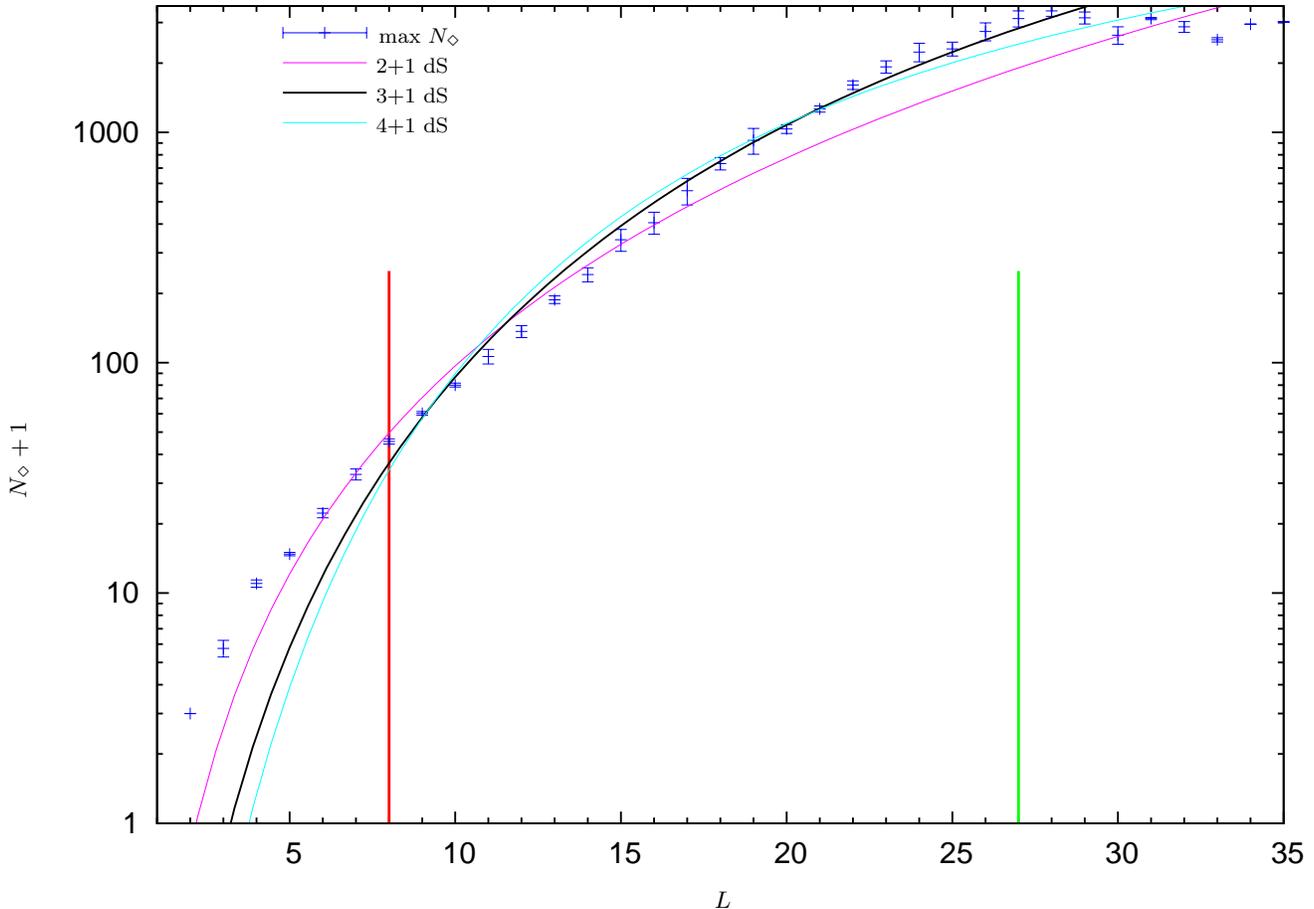}
\caption{A plot of the maximum values of $\nd$ as a function of $L$ for
  $p=0.0002$ and $N=30000$, in log scale, along with best fit curves for \ds space of three
  different dimensions.
Again the overall best fit is achieved from the curve for 3+1
  dimensions. 
}
\label{p0.0002.fig}
\end{figure}

\begin{figure}[htbp]
\psfrag{data2}{\hspace{25mm}\footnotesize max $\nd$}
\psfrag{f2}{\hspace{18mm}\footnotesize 2+1 dS}
\psfrag{f3}{\hspace{18mm}\footnotesize 3+1 dS}
\psfrag{f1}{\hspace{18mm}\footnotesize 1+1 dS}
\psfrag{Ndiamond+1}{$\nd+1$}
\psfrag{L}{$L$}
\includegraphics[width=\textwidth]{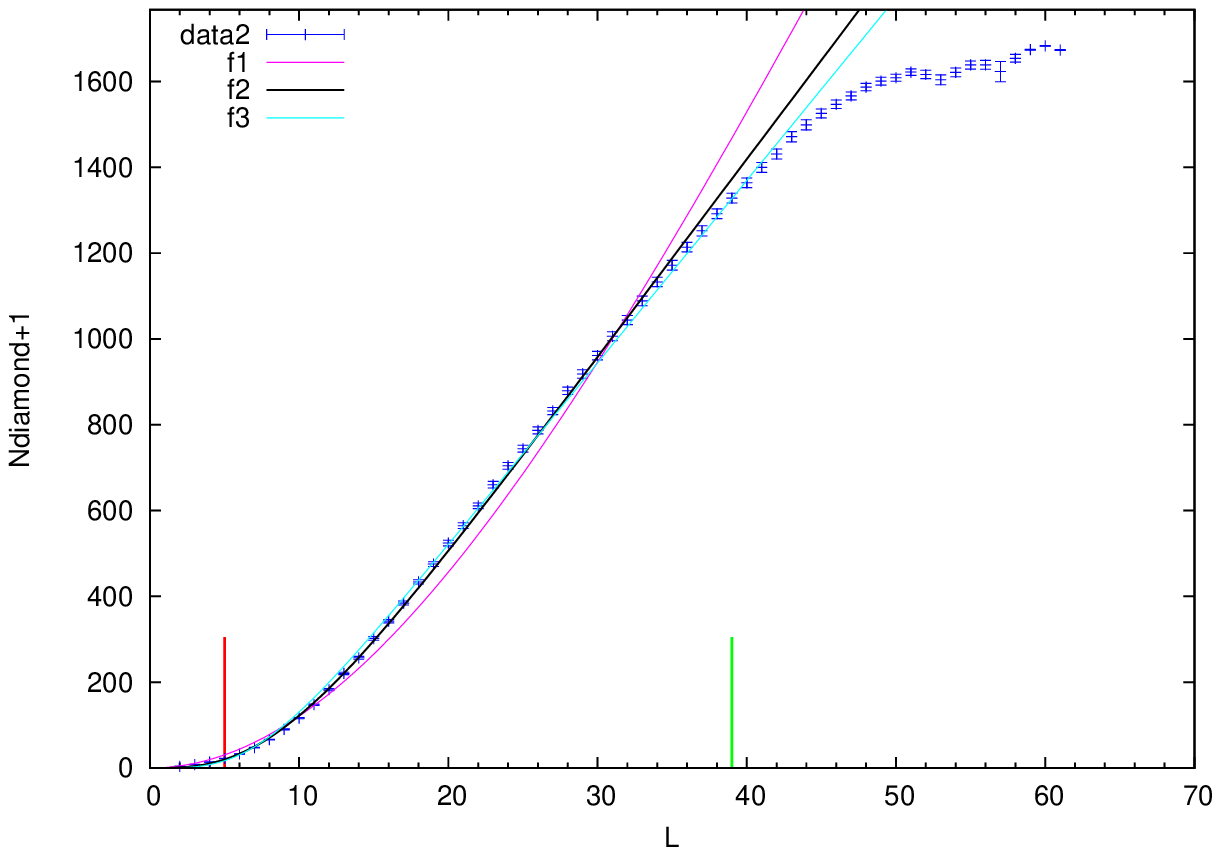}
\caption{A plot of the maximum values of $\nd$ as a function of $L$ for
  $p=0.01$ and $N=2000$, along with best fit curves for \ds space of three
  different dimensions.
Here the overall best fit is achieved from the curve for 2+1
  dimensions.}
\label{p0.01.fig}
\end{figure}

Figure \ref{p0.001.fig} contains the results from our largest data set
(largest number of causal sets generated with those parameters).  The plot
for our smallest value of $p$ is shown in figure \ref{p0.0001.fig}.
There the range of $L$ values available for the fit is smaller, because one
needs a very large causal set to get large chains with such a small $p$.  The
curve for 3+1 \ds continues to make an excellent fit, and better than curves
for different dimensions of \ds space.
Figure \ref{p0.0002.fig} portrays another example of the fits, this time in
log scale.
Our final example, figure \ref{p0.01.fig}, comes from a larger value for $p$.
Here the universe is quite small, with a radius of curvature of just $\sim
4$ in fundamental units.  The best fit dimension is only 2+1 for this tiny
universe.

\begin{figure}
\psfrag{Ndiamond+1}{$\nd$}
\psfrag{L}{$L$}
\psfrag{f2}{\hspace{18mm}\footnotesize 2+1 dS}
\psfrag{f3}{\hspace{18mm}\footnotesize 3+1 dS}
\psfrag{f4}{\hspace{18mm}\footnotesize 4+1 dS}
\psfrag{f5}{\hspace{18mm}\footnotesize 5+1 dS}
\psfrag{f6}{\hspace{18mm}\footnotesize 6+1 dS}
\psfrag{data2}{\hspace{25mm}\footnotesize max $\nd$}
\psfrag{data6}{\hspace{25mm}\footnotesize mean $\nd$}
\includegraphics[width=\textwidth]{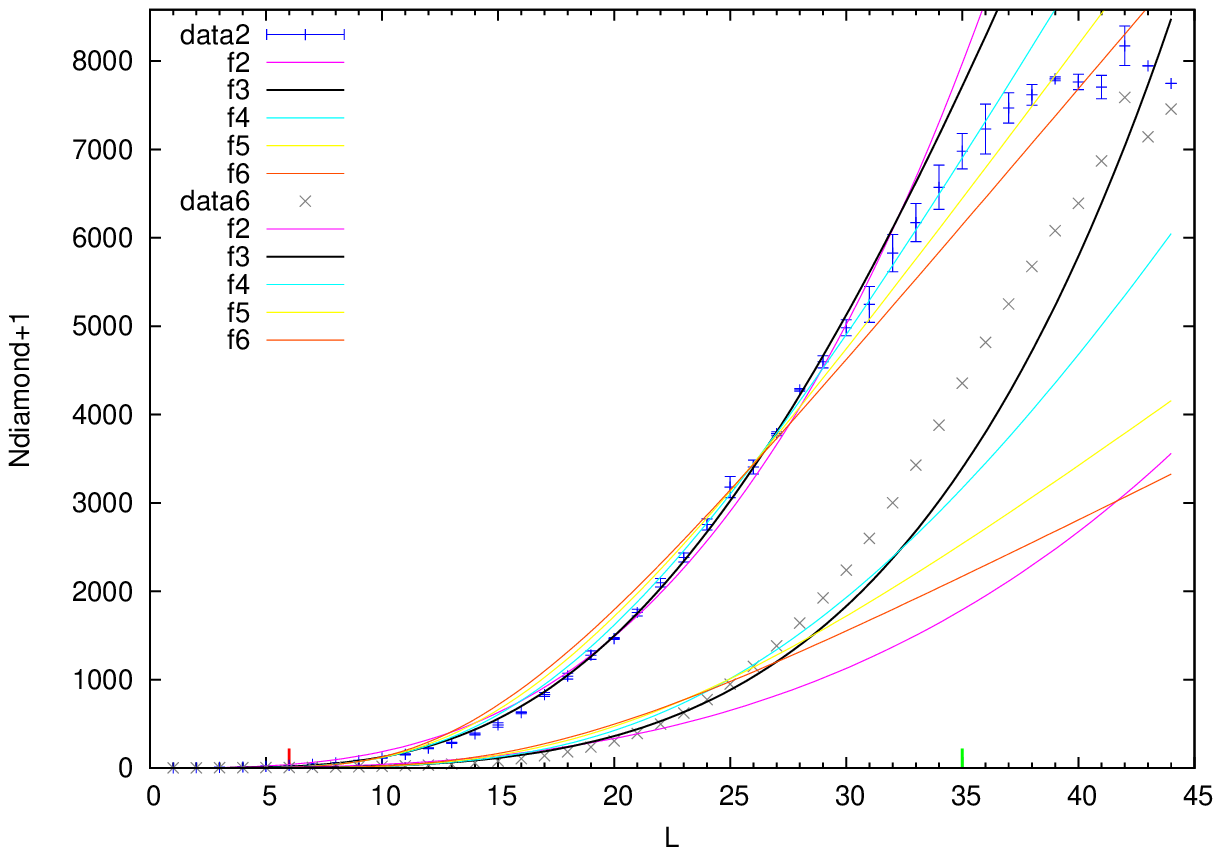}
\caption{Comparing behavior of the mean $\nd$ versus the maximum.  The
  results come from causal sets generated with $p=0.0008$, $N=15000$.}
\label{mean_vs_max.fig}
\end{figure}

Before concluding, we return attention to the issue of fitting the mean
vs.\ the maximum $\nd$.  In figure \ref{mean_vs_max.fig} we directly compare the
two on the same data set, generated from four $N=15000$ $p=0.0008$ causal sets.
It is clear that the maximums fit the 3+1 \ds form quite well, while the mean
values do not.  We omit the errors in the means, because they are extremely
small for the small intervals, and strongly bias the fits away from the data
at larger $L$.  In any event the fit is poor for the means, for any dimension
(though again 3+1 fits the best there as well).

\section{Summary and Conclusions}
\label{conclusions.sec}

After motivating the study of originary percolation as an appropriate
dynamical model for the early universe of Causet Set Theory, at least within
the context of classical sequential growth models, we explored a number of
indications that it yields an exponentially expanding universe.  In
particular, for $p\ll 1$, we saw that, after a post, the universe begins with
a random tree era, followed by a period of \ds-like exponential expansion.
More specifically,
for $p\ll 1$ and $N\gg p^{-1}$,
the largest intervals in the post-tree era resemble 
de Sitter spacetime insofar as spacetime
volume as a function of proper time is concerned.
Furthermore, the expression that best fits the data has $D=3$ for a
significant range of $p$ (at least one and a half decades).  If $D$ continues
to vary monotonically with $p$, then our results are compatible with $D=3$
all the way down to physically realistic values, say $10^{-84}$ as needed to
explain the initial large size of the universe \cite{causetcosmo}.
Does this indicate how the observed number of spatial dimensions will emerge
dynamically from quantum gravity?
One must wait for the full quantum theory to be sure, but the dynamical
appearance of 3+1 dimensions, without being put into the theory in any way,
is intriguing.

~\\
There are many arguments that motivate the 
assumption of a discrete structure for our universe 
at the most fundamental level, and Causal Sets are a very simple 
and clearly defined theory that does just that. Some of these arguments
which are mostly philosophical in nature
are powerful and have been around for a very long time 
but the lack of any observational effects of discreteness has left
the idea as a beautiful orphan that few want to adopt.
It has only been recently that through arguments that derive life
from Causal Set theory have we been able to predict some observational 
effects of discreteness as well. Fluctuations in the cosmological 
``constant'' are one such prediction. Now we have shown that the 
universe generated by (many of) the {\it CSG} models not only exhibits
some very desirable cosmological properties but may help solve some of the 
toughest problems of the standard cosmology, such as:

\begin{itemize}

\item The standard model of cosmology does not tell us from where the universe
comes. In fact, if the theory of General Relativity is supposed to
be valid all the way to time $t=0$, the universe ends up in a singularity,
where not only the physical laws do not apply but it is impossible 
to get any information from $t<0$.  Thus it is impossible to 
know what happens ``before'' the singularity. On the other hand if Causal
Set cosmology is taken seriously, one still has a ``beginning'' 
or a big bang in the model but the singularity is not a problem anymore. 
The post is like any other element in the theory and thus discreteness
can ``resolve'' the singularities. In fact, the same post is the big crunch
singularity of the previous cycle of the universe. One can in 
principle calculate the probability of post occurrence 
in any of the {\it CSG} models, and it is easy to see what happens
to the universe before and after the post is formed.

\item Every time the universe collapses (to a post) and then bounces back, the 
effective behavior of the expansion can be described as if the 
whole causal set started with that post with renormalized 
coupling constants. 
Since the percolation dynamics is an attractive fixed point under
this renormalization flow in 
the space of CSG models that have posts, one may start the universe
generically in any of these models, and it eventually will end up arbitrarily
close to percolation. 
This makes percolation the natural candidate for
the study and also guarantees the results are free of 
any kind of fine tuning in the space of models.

\item The universe in the percolation model has two clearly separable eras
early on. The first of these resembles a random tree, where the 
spatial volume of the universe
increases exponentially with the ``cosmological time''. As the universe 
accumulates $1/p$ elements after the post, 
where $p$ is the parameter of the 
percolation, it enters a de Sitter like phase.

\item One of the most unsettling problems of the standard cosmology is
the fact that the universe appears very homogeneous on large scales ---
something that can be seen directly in the Cosmic Microwave 
Background temperature isotropy. 
The percolation
universe as it emerges from
its early phase is very homogeneous in the sense that any neighbourhood
looks like any other.  Every element has the same sort of past and future and
the same number of nearest neighbours. Thus the model has a very strong 
potential for solving the homogeneity problem as it naturally favours a
homogeneity in the initial conditions. This is particularly
true if the matter is generated by the structure in the Causal
set itself. On the other hand, if we put external degrees of freedom
on the Causal set, it may happen that even 
if we start with different initial conditions for these 
degrees of freedom, the de Sitter like expansion gets rid of this
inhomogeneity. 
Of course there are 
random fluctuations that cause deviations away from homogeneity. These
fluctuations might prove helpful in solving another extremely
important puzzle in the early universe, namely the origin of 
density perturbations that seed the late time structure formation.

\item Another puzzle is the large size of the universe compared to, say,
the Planck length, when the universe is still very young, say, 
$O(100)$ Planck times old.
This is related to both the horizon problem and the flatness puzzle. Models
with percolation dynamics 
naturally generate a large size of the universe. If we start a percolation
model with parameter $p$ the spatial volume becomes of the order of 
$p^{-1}$
within $\ln{p^{-1}}$ time steps. Depending on how small $p$ is, the 
universe can be made arbitrarily large. Since cosmic renormalization provides
a mechanism which
can drive the effective value of $p$ to arbitrarily small values if one waits long enough, there is no
fine tuning involved.

\end{itemize}

It may be the case that the quantum mechanism which drives the cosmological
constant to zero \cite{sor90}
is the same mechanism which causes a smooth continuum to
emerge from the discrete partial order.  In this case one may not be so
surprised that the CSG models do not lead to smooth continuum-like manifolds.
However, it is possible that they capture some new physics at cosmological
scales, given their discrete nature.  Here we have demonstrated that CSG
models are easily capable of describing a rapidly expanding universe which is
much like our own, at least at the largest scales.  Could
the locally Minkowskian light cone structure of continuum spacetime be an
effect which arises only at an intermediate scale, much larger than the
discreteness scale, and thus is not a good description of our universe until
after an initial period of \ds like expansion?

~\\
\noindent\textbf{\large Acknowledgements}\\
We are extremely grateful to Niayesh Ashfordi for numerous discussions on the
content of this paper.  We also thank Rafael Sorkin, Raissa D'Souza, and Tim Koslowski for
illuminating discussions.
This research was supported by the Perimeter Institute for
Theoretical Physics.  Research at Perimeter Institute is supported by the
Government of Canada through Industry Canada and by the Province of Ontario
through the Ministry of Research \& Innovation.
The numerical results were made possible in part by the facilities of the
Shared Hierarchical Academic Research Computing Network
(SHARCNET:www.sharcnet.ca).
We also thank Yaakoub El Khamra for providing computational resources and
advice.

\end{document}